\documentclass[letterpaper]{article} 
\usepackage{aaai2026}  
\usepackage{times}  
\usepackage{helvet}  
\usepackage{courier}  
\usepackage[hyphens]{url}  
\usepackage{graphicx} 
\usepackage{natbib}  
\usepackage{caption} 
\usepackage{algorithm}
\usepackage{algorithmic}
\usepackage{booktabs}

\usepackage{xcolor}

\usepackage{newfloat}
\usepackage{listings}
\DeclareCaptionStyle{ruled}{labelfont=normalfont,labelsep=colon,strut=off} 
\floatstyle{ruled}
\newfloat{listing}{tb}{lst}{}
\floatname{listing}{Listing}

\usepackage{amsmath}

\title{How Early Adopters Used Generative AI Worldwide: \\ Variation by Country Income and Language}

\author{
    Madeleine I. G. Daepp\textsuperscript{\rm 1},
    Isaac Slaughter\textsuperscript{\rm 2}
}

\affiliations{
    
    \textsuperscript{\rm 1}Microsoft Research\\
    \textsuperscript{\rm 2}University of Washington\\

     mdaepp@microsoft.com, is28@uw.edu
    
}

\nocopyright
\begin{document}

\maketitle

\begin{abstract}

AI is being used by people globally, but not everyone is using it in the same ways. Using a large-scale dataset of anonymized, de-identified, and privacy-scrubbed interactions with a widely available and free AI chatbot, we empirically characterize differences in early adopters' usage across countries. Schooling is the most common domain of use in most countries, particularly low-income countries, with a strong inverse association evident between schooling and country-level GDP. Leisure-related use, by contrast, is positively associated with country-level income. Language, we find, also shapes use: English-language interactions are overrepresented in places where the predominant languages were not well-served by existing models during the period of the study. Improving performance across languages may be a key factor, our work suggests, in whether this technology expands digital divides or enables leapfrogging.

\end{abstract}

\section{Introduction}

Generative AI has seen rapid uptake around the world, but its diffusion has been uneven~\citep{liu_who_2024, microsoft2026globalai}. In March 2024, low-income countries accounted for less than 1\% of visits to generative AI websites~\cite{liu_who_2024}. By the second half of 2025,  adoption was growing twice as fast in the Global North as in the Global South~\cite{microsoftAIAdoption2025}. Countries whose populations speak languages commonly represented in AI training data sets have adopted generative AI at approximately twice the rate of countries whose populations do not \cite{misra2025ai}. Taken together, these patterns suggest that global AI adoption is consistent with a classic ``digital divide`` in which prior disadvantages are reproduced through differential access to new technologies~\cite{hargittai2003digital, van_dijk_digital_2006}.

Disparities in access, however, are only the first step in understanding how a technology's impact will be distributed. Digital divides can persist even after access is achieved, through differences in how technologies are used and in the skills needed to use them effectively---what has been termed the ``second-level'' digital divide~\cite{hargittai2003digital}. In studies of early internet adoption, for example, more educated users were more likely to engage in information seeking, email, and news, while less educated users were more likely to spend time on social media~\cite{goel2012does, buchi2016modeling}. Language also shaped use: Early studies of Internet use in Egypt, for example, found that the absence of content in users' primary language constrained the range of activities they could meaningfully engage in online~\cite{warschauer_technology_2003}. Even when access barriers are reduced, that is, divides can persist through differences in use and linguistic accessibility.

It is important to understand whether similar usage divides characterized the early adoption of AI, because early patterns of use shape who benefits from these systems. Despite a growing body of work examining how AI is used around the world~\cite{chatterji_how_2025, appel_anthropic_2025, costa2025s, yang2025adoption}, as well as studies examining divides in access and adoption within and between countries~\cite{daepp_emerging_2025, misra_measuring_2025, liu_who_2024}, there remains a need for large-scale and cross-national characterization of the needs for which people use generative AI around the world. 

In this paper, we address this gap by studying two questions about global differences in early generative AI use:
\begin{enumerate}
    \item \textbf{RQ1: How do the purposes for which early adopters use generative AI vary across countries?} To answer this question, we develop and validate a novel use-case classifier based on a labor-economic taxonomy of how people allocate their time \cite{ramey_century_2009}. Schooling is the most prominent use of AI across countries, with more usage for schooling the lower income the country; by contrast, use of AI for leisure is positively associated with country-level income.
    \item \textbf{RQ2: How does language use in interactions with generative AI vary across countries?} We answer this question by comparing language use in chatbot interactions to country-level language distributions and model performance benchmarks. English is strongly overrepresented among AI users in Asian and African countries, with languages for which models perform poorly particularly underrepresented in in-the-wild chatbot use.
\end{enumerate}

Our study leverages a unique dataset of deidentified, anonymized, and privacy-scrubbed conversations from one of the first free, globally available AI chat interfaces. We construct a country-stratified sample comprising 54,841 users and 686,722 conversations across 227 unique countries, spanning a six-month period from April to September to 2024.

 Through this work, we make the following contributions:
\begin{itemize}
    \item We provide a large-scale, cross-national analysis of how generative AI is used globally, based on early adopters' real-world interactions with AI.
    \item We introduce and validate a novel use-case classifier grounded in prior research on time use in the labor economics literature.
    \item We show that patterns of AI use vary systematically with national income:  use for schooling is most prevalent in lower-income countries, while leisure-related use increases with income.
    \item We show that users disproportionately rely on English when local languages are poorly supported by AI systems, providing one of the first in-the-wild studies of how language limitations shape global patterns of AI use.
\end{itemize}

\section{Related Work}

\subsection{Digital Divides in AI Access}

As computers and the Internet spread in the late 1990s, scholars adopted the phrase ``digital divides'' to describe inequalities in access to and successful use of digital technologies \cite{hargittai2003digital}. Early research documented widespread ``first-level'' divides across geographies or demographics: disparities in who had physical or material access to the Internet and computers \cite{van_dijk_digital_2006,van_deursen_first-level_2019}. Later work on ``second-level'' divides showed that even once people had the means to access a technology, disparities remained in the frequency with which they did so, the uses they had for it, and the skills they developed for its use \cite{van_deursen_digital_2014, redmiles2018net}.\footnote{Scholars later adopted the phrase ``third-level divides'' to refer directly to disparities in the harms and benefits that people gain from a technology \cite{weiConceptualizingTestingSocial2011}.} 

These disparities have been observed in the adoption of AI, as well.  Adoption is higher, for example, among younger versus older and more versus less educated people \cite{kacperskiCharacteristicsChatGPTUsers2025,draxlerGenderAgeTechnology2023,bick2026rapid,leeDigitalDivideGenerative2026,chatterji_how_2025}. There is also strong evidence of geographic divides, with early research documenting urban hotspots versus left-behind rural regions in the United States \cite{daepp_emerging_2025} and across OECD countries more broadly \cite{muroGeographyAIWhich2025,oecdJobCreationLocal2024}. High-income countries have the highest rates of usage \cite{chatterji_how_2025, misra_measuring_2025} and, although adoption has increased considerably in low- and middle-income countries \cite{yang2025adoption,appel_anthropic_2025}, there is a persistent and widening adoption gap between Global North and Global South countries \cite{microsoft2026globalai}.

While the first-level divide in access and adoption remains an important challenge, the drivers of second-level divides in AI usage are less clear. One factor may be the difference in local economies and thus in user needs, because AI adoption is highly uneven across different occupations. Knowledge workers were significantly overrepresented among early AI adopters \cite{bick2026rapid,gillespie_trust_2025}, and early adopters of agentic AI tools are disproportionately from knowledge-intensive occupations \cite{yang2025adoption, handa2025interviewer}. Consistent with the idea that differences in local economies could lead to differences in the uses for and usefulness of AI, \citet{huang2026interviewer} finds that the purposes people envision for AI vary systematically by economic context. Users in lower-income regions are more likely to envision AI as a tool for entrepreneurship and learning, while those in wealthier nations are more likely to envision AI as a tool for life management and personal transformation. Notably, differences in domains of use were an important feature of early second-level digital divides research, which showed that more advantaged people tended to spend time in productivity-enhancing applications (email, news) while less advantaged users users were more likely to use the internet for distraction (social media, entertainment) \cite{goel2012does, buchi2016modeling, smirnov2018predicting}. There is a need for further research to understand whether this same pattern emerges in AI adoption, because the extent to which users spend time with generative AI for human capital accumulation—--such as work, schooling, or household production—--versus for leisure could lead to long-term differences in economic trajectories.

\subsection{How and Why People Use Gen AI}

To identify the domains for which AI is used, researchers have proposed a number of different taxonomies and classification methodologies. One line of research draws on the well-established literature on user intent in information retrieval, adapting taxonomies originally developed for search log data to chat logs \cite{zamaniConversationalInformationSeeking2023,shahUsingLargeLanguage2025}. These studies observe that users often turn to AI for guidance or information---finding, for example, that for practical guidance, information seeking, and writing account for approximately 4 in 5 ChatGPT messages \cite{chatterji_how_2025}, or that learning and communicating were among the most common goals for users of Bing Copilot \cite{tomlinson_working_2025}. More recent literature also argues that chatbot interactions span not only instrumental needs like information seeking, synthesis, and content creation, but also that conversations can provide users with social interaction and ``meta-conversation,'' such as concerning a chatbot's capabilities \cite{shelbyTaxonomyUserNeeds2025}. Notably, however, conversations with AI are considerably more complex than search \cite{shahUsingLargeLanguage2025}, and there is a need for the development and validation of classifiers that look beyond information retrieval and economic tasks literature to better characterize how generative AI is fitting into people's lives. 

A second common approach is thus to map AI's use to economic categories. \citet{eloundouGPTsAreGPTs2024}, for example, develop a taxonomy based on the U.S. Department of Labor’s O*NET framework, which organizes work into occupations, skills, and detailed work activities. Researchers have since used this approach to examine the occupational role the chatbot is playing \cite{handa_which_2025}, the occupational task to which individual message relates to \cite{chatterji_how_2025}, the conversation in general \cite{appel_anthropic_2025}, or the activities that either the chatbot or human user are carrying out \cite{tomlinson_working_2025}. These works have consistently found information-related tasks to be heavily represented in conversations with AI, for example that conversations typically pertain to gathering information, coding, writing, and making decisions \cite{tomlinson_working_2025,chatterji_how_2025,handa_which_2025}. \citet{tomlinson_working_2025} further find high applicability of AI to knowledge work tasks---a result consistent with the growing field and experimental evidence that generative AI's use enhances efficiency in knowledge work domains \cite{dillon_shifting_2025,brynjolfssonGenerativeAIWork2025,cuiEffectsGenerativeAI2026}. Knowledge work is disproportionately concentrated in developed and developing economies, however, and there is a need for taxonomies that are also relevant to the low-income economies in which household production and other forms of labor are more common \cite{bjorkegren2026poorestai}.

\subsection{Language and AI Adoption}

Language contributes to cross-country differences in AI adoption, use, and usefulness. While chatbots theoretically allow users to interact in the language of their choosing, performance varies widely \cite{joshiStateFateLinguistic2020}. Large language models work best, benchmarking studies show, in the ``high-resource'' languages commonly found online, and performance across a variety of tasks degrades considerably for ``low-resource'' languages that are less prevalent online and thus less represented in training data \cite{xuan_mmlu-prox_2025, singhGlobalMMLUUnderstanding2025, huangBenchMAXComprehensiveMultilingual2025a, romanou2025include,vayani2025all}. These performance differentials likely influence adoption: countries where low-resource languages predominate have average adoption rates that are 20\% lower than predicted after accounting for demographic factors, economic advantage, or internet access \cite{misra2025ai}. Most existing work relies on the role of language in shaping chat interactions relies, however, on lab-based benchmarks or convenience samples---such as interactions with specialized interfaces set up for research purposes~\cite{zhaowildchat,chiangChatbotArenaOpen2024,zhengLMSYSChat1MLargeScaleRealWorld2023}, with little research actually examining language use in-the-wild for major AI products. That is, while linguistic disparities in model performance are well-documented, we lack large-scale and real-world evidence on the languages in which real users actually interact with generative AI systems. Our study seeks to fill this gap by examining the languages in which early adopters used AI in a large-scale and cross-national sample. Through this work, we contribute to the emerging literature on digital divides in access to and use of generative AI globally.

\section{Data}\label{sec:data}

\begin{table}[t]
\centering
\begin{tabular}{p{2.1cm} p{5.4cm}}
\hline
\textbf{Category} & \textbf{Definition} \\
\hline
\textsc{Schooling} &
Degree-based learning or assignment-like work. Includes conversations where user asks formal or structured questions as might appear on a school assignment. \\
\textsc{Market Work} &
Paid work, job search, or projects that improve employability. Includes non-school projects that improve the user's earning potential (e.g., personal websites, game development), work at non-profits, and work as a teacher or other educator.
 \\
\textsc{Household Production} &
Household tasks substitutable by market services, such as cooking, cleaning, childcare, home or vehicle maintenance, personal device maintenance, personal finance, event planning, or obtaining housing. \\
\textsc{Personal Care} &
Assistance with care of the self, including health, mental health, nutrition, sleep, fitness, grooming, or beauty-related activities. \\
\textsc{Leisure} &
Entertainment, hobbies, socialization, play, or informal self-learning with no clear external stakes or deadlines. \\
\hline
\end{tabular}
\caption{Conversation Purpose Taxonomy, Adapted from \citet{ramey_century_2009} The full classification prompt used for annotation is provided in the Appendix.}
\label{tab:ai_use_taxonomy}
\end{table}

Our work relies on a dataset of de-identified, anonymized, and privacy-protected conversations from users of a widely available AI chat system, Microsoft Bing Copilot. 
All conversations were anonymized using in-house privacy filters to remove personal information, such as names, contact information, and addresses. Users are identified by irreversibly hashed IDs, preventing the possibility of re-identification, and no personally identifiable or sensitive information was visible at any stage of the analysis. All data handling and analysis procedures were reviewed by Microsoft's appropriate ethics and privacy teams to ensure compliance with the company’s 
responsible AI principles and the terms of use for users of the free Bing Copilot product in 2024, which allowed use for the purposes of research. Data access was limited to the authors of this work, and all underlying user data were deleted after the construction, validation, and application of classifiers, with the analyses relying only on the outputs of these classifiers.

To understand how early adopters' usage varied around the world, we took a stratified random sample of up to 250 users per country from the six-month period beginning April 1st, 2024 and ending September 30th, 2024. Users active during our collection period were assigned to the country in which their sessions most frequently occurred, with ties broken by the location of their final session. For each selected user, we randomly sampled up to 20 conversations. If a user had fewer than 20 conversations, we included all available conversations. To ensure our sample is representative of ``early adopters'' or people who repeatedly used AI chat, and not those who conducted one-off tests, we excluded users with fewer than 5 conversations. We also excluded any countries for which there were fewer than 50 unique users to protect user privacy. Our final analytic dataset comprises 54,841 users and 686,722 conversations across 227 unique countries. 

We reserve additional development and validation sets for use in constructing our classifier, randomly sampling users across countries in each World Bank income group and retaining those with at least five sampled conversations. After applying this criterion, the development set contained 50 users (590 conversations) and the validation set contained 58 users (756 conversations). In total, these datasets comprised 108 users and 1,346 conversations spanning 82 unique countries and all four income groups.

Finally, we augmented this sample with several country-level data sources. We calculated country-level income using Gross National Income (GNI), Gross Domestic Product (GDP), and population estimates from the World Bank's World Development Indicators (WDI) for 2024. To estimate the total number of English speakers in each country, we used data from the Unicode Common Language Data Repository (CLDR) and the LinguaMeta open-source language metadata repository, following \citet{ritchie_linguameta_2024}.

\section{Methods}

To better understand how early adopters around the world spent time with chatbots, we developed a classifier based on a validated taxonomy from the labor economics literature \cite{ramey_century_2009}. Two researchers labeled the development data, discussing and resolving disagreements, producing a gold-standard set of labels. We then used this labeled development set to iteratively refine the classifier. Initially, we adapted the prompts from \citet{handa_which_2025}, \citet{tomlinson_working_2025} and \citet{phang_investigating_2025} to the domain of use task. We observed substantially improved performance when the classifier was conditioned on multiple conversations from the same user before labeling individual interactions. As a result, we constructed the classifier in two stages as follows. First, we applied an LLM to produce a concise summary of user context from the complete set of conversations; second, we input both that summary and a single conversation, and prompted an LLM to apply a domain of use classification to the single conversation. Our final prompts also leveraged procedural generation, instructing the model to utilize the \texttt{<quotes>} tag to identify direct quotes relevant to the labeling task at hand and \texttt{<thinking>} tag to aid in forming a response before one is provided, and structured output. The complete text of our final prompts can be found in the Appendix. 

To validate our classifier, we again had the two internal human reviewers, who were blind to the AI labels, independently label the conversations in the validation set. Comparing two human and one AI reviewer, we obtain a Fleiss's $\kappa$ of 0.605 and Krippendorf's $\alpha$ of 0.603. These levels of reliability, while moderate, are comparable or exceed those achieved in similar prior work \cite{chatterji_how_2025, tomlinson_working_2025}. The AI classifier also achieves substantial agreement compared to human consensus ($\kappa$ = 0.743), suggesting that disagreements represent genuine ambiguity in the labeling task rather than systematic error; results compared to consensus, moreover, are nearly identical across the development and validation sets. Finally, the AI-H1 pair achieves higher pairwise agreement than the human-human pair (73.8\% vs. 69.8\%), indicating that the AI performs at least as well as a second human coder. Full details on classifier development and inter-rater reliability can be found in the Appendix.

Our final classifier applies English-language prompts to multilingual data, an approach consistent with prior work \cite{handa_which_2025, tomlinson_working_2025}. For the purposes of labeling the development and validation sets, we translated any non-English conversations to English, using gpt-4o-mini-0718 to carry out a multi-pass translation strategy shown to perform comparably with more complex translation techniques \cite{wu2025please}. However, to ensure our inter-rater reliability metrics are consistent with how the classifier would actually be used, we compared our labels against those produced by running the final classifier with the original multilingual data. Our translation prompts are included in the Appendix.  

Finally, we identified the language contained in each conversation using GlotLID, a language identification model that can identify 1665 languages and achieves an average F1 greater than 0.9 when evaluated on a diverse set of multilingual benchmarks \cite{kargaran_glotlid_2023}. 

All classification, processing, and translation was carried out with an internally hosted or local model in a closed compute environment, in keeping with privacy and IRB requirements, to ensure no user data was processed externally.

\section{Analysis and Results}

\subsection{Domain of Use}

\begin{figure*}[t]
\centering
\includegraphics[width=0.9\textwidth]{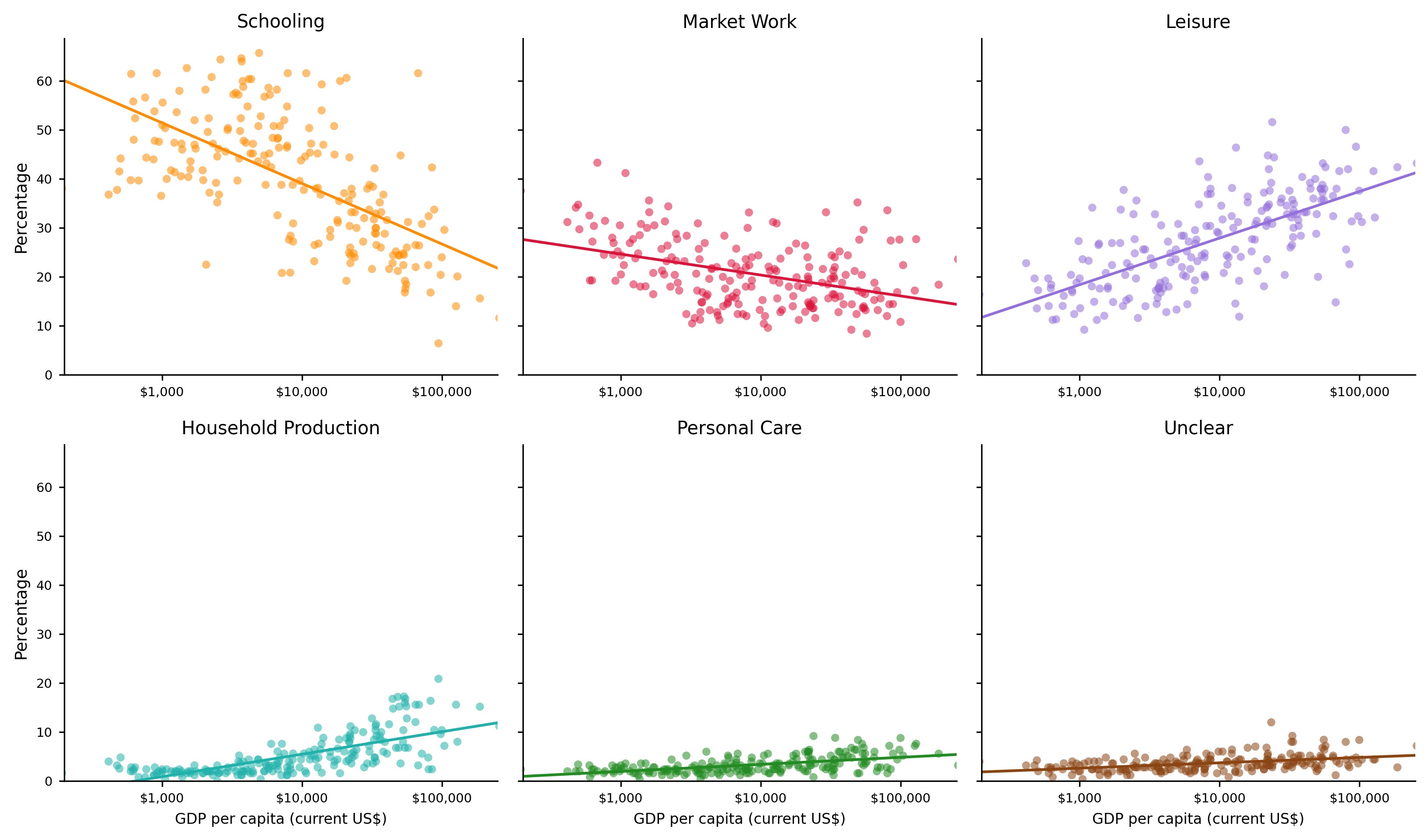}
\caption{Frequency of Conversation Purposes Relative to Country GDP. Points show the fraction of early adopters in a given country who predominantly had conversations in a given domain.}\label{fig2:domains}
\end{figure*}

We first examine the domains in which early adopters used AI (RQ1). Figure~\ref{fig2:domains} shows how early adopters' primary domains vary as a function of GDP per capita (log scale). Schooling is a major domain of use and is the modal domain for a majority of users in 66.7\% of countries. There is an inverse relationship, however, between usage for schooling and logged country-level income (Spearman's $\rho = -0.64$). Market work  similarly shows an inverse association ($\rho = -0.38$), but usage for leisure is strongly positively associated with log GDP ($\rho = 0.69$). Although household production and personal care together constitute primary domains for fewer than 10\% of users across countries, there is some evidence of higher usage for these domains in higher income countries ($r = 0.76$ and $0.52$, respectively). Domains classified as \textit{unclear} account for a relatively small share of usage, indicating that most conversations can be meaningfully categorized within the taxonomy. All reported associations are statistically significant based on two-sided Spearman correlation tests with the Benjamini-Hochberg false discovery rate correction for multiple testing (p $<$ 0.01) \cite{benjamini1995controlling}.

\begin{figure}[t]
\centering
\includegraphics[width=0.9\columnwidth]{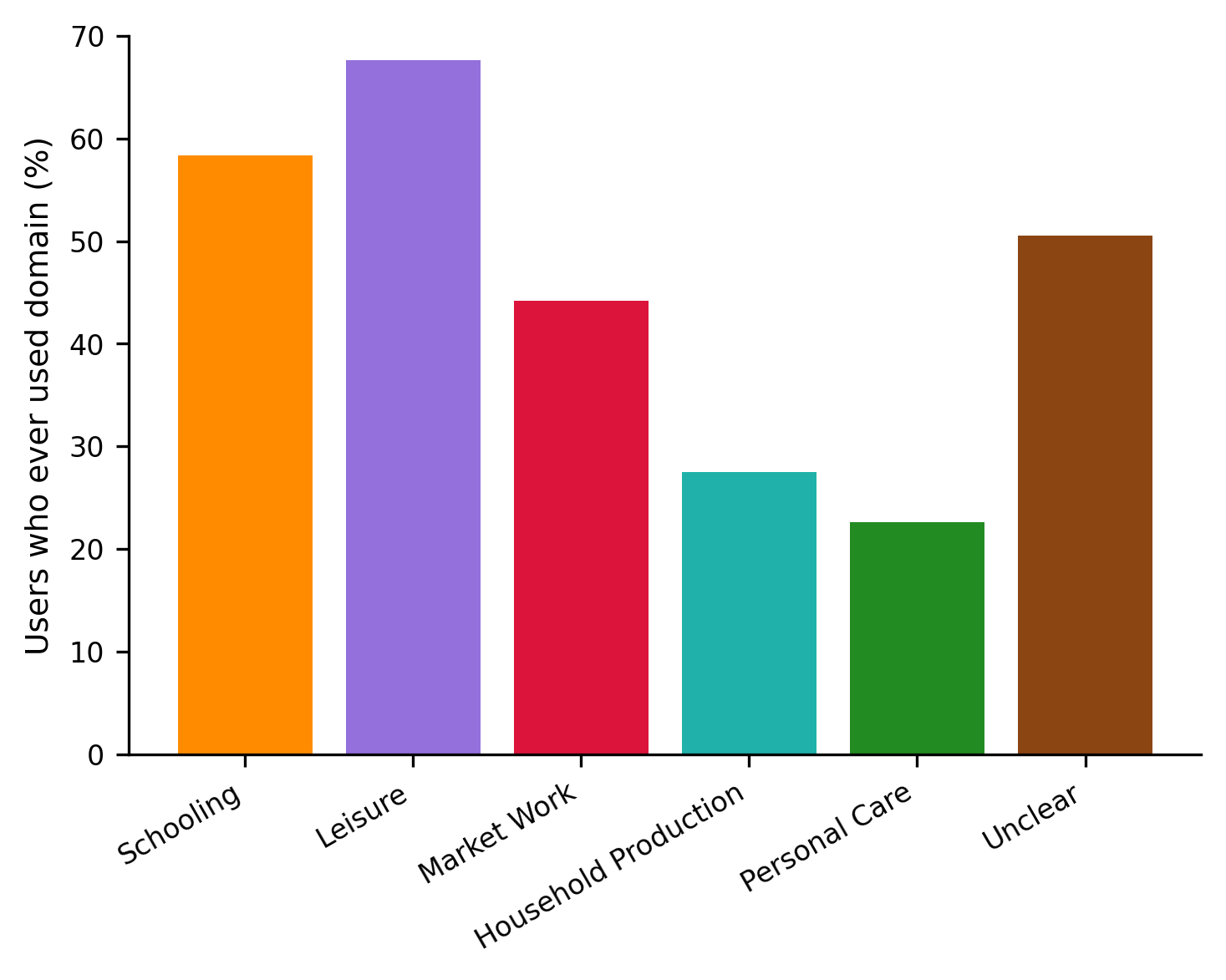}
\caption{Fraction of early adopters in sampled data with at least one conversation in each domain.}\label{fig2:bar}
\end{figure}

\begin{figure}[t]
\centering
\includegraphics[width=0.95\columnwidth]{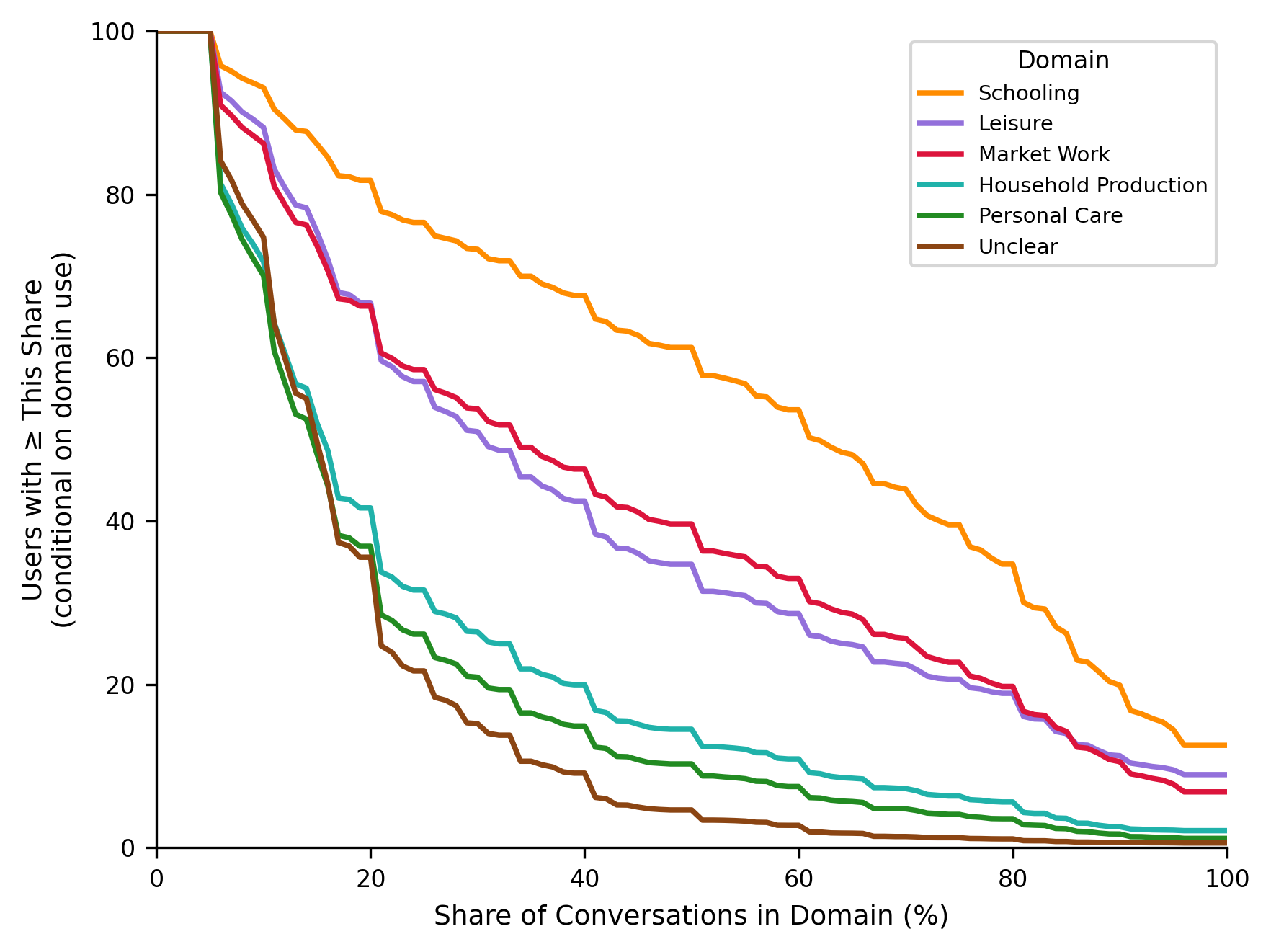}
\caption{Concentration of domain use. Lines show the complementary cumulative distribution function of the share of conversations devoted to each domain, conditional on early adopters having engaged with that domain at least once.}\label{fig3:cdf}
\end{figure}

Most users in the sampled data have at least one conversation that is classified as leisure (67.6\% of users) and at least one conversation classified as schooling (58.3\% of users) (Figure~\ref{fig2:bar}). Figure~\ref{fig3:cdf} shows the complementary cumulative distribution function (CCDF) of the share of conversations devoted to a given domain among users who ever had a conversation classified in that domain. Notably, the CCDF for schooling is convex, indicating that users who ever use the chatbot for schooling tend to devote a large share of their conversations to that domain. Among users who ever use the chatbot for schooling, 57.8\% devote the majority of their conversations to schooling. For other domains, the curves are concave, indicating more dispersed usage. Among users who ever use the chatbot for conversations in the leisure domain, for example, just 31.4\% have a majority of their conversations in that domain. Appendix Figure~\ref{appxfig2:cdf} shows that these patterns are consistent across country-income levels.

\begin{figure*}[htbp]
\centering
\includegraphics[width=0.95\textwidth]{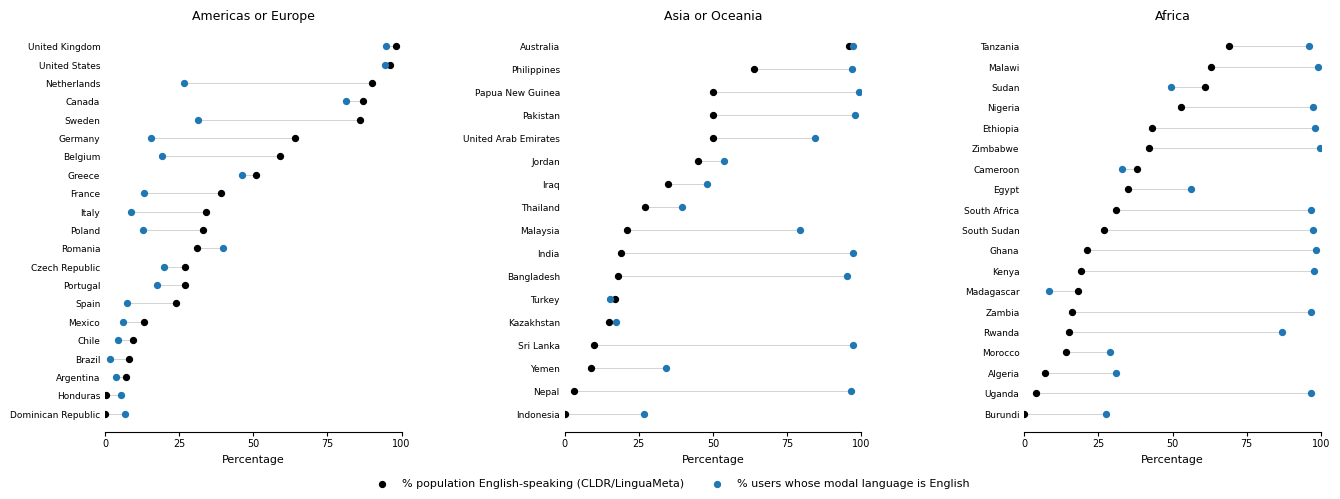}
\caption{English language use in AI Conversations relative to estimated English prevalence in the general population, by country. Blue points show the fraction of users in a given country whose predominant usage occurs in English; black points show the fraction of the overall population estimated to have English competency (max of Unicode CLDR and LinguaMeta estimates \cite{ritchie_linguameta_2024}). \footnote{Statistics are calculated for frequent users only (More than 5 conversations over the study period) using a country-stratified sample. Only countries with population $\geq$ 10M and $\geq$ 50 sampled users are included.}}\label{fig5:language}
\end{figure*}

\subsection{Language Use}

We next examine the languages in which early adopters prompted the chatbot (RQ2). Figure~\ref{fig5:language} shows the fraction of users in the sampled data for whom English was the modal language for AI usage relative to estimated English-language competency at the population level. Note that these estimates are noisy \cite{ritchie_linguameta_2024, kargaran_glotlid_2023}; nevertheless, a noteworthy pattern emerges. In particular, in Europe and the Americas, the estimated fraction of the population with competency in English generally exceeds the fraction of users for whom English is the modal language. By contrast, English disproportionately functions as the lingua franca of AI usage in Asia/Oceania and Africa relative to estimates of population competency in the language.

\begin{figure}[t]
\centering
\includegraphics[width=\columnwidth]{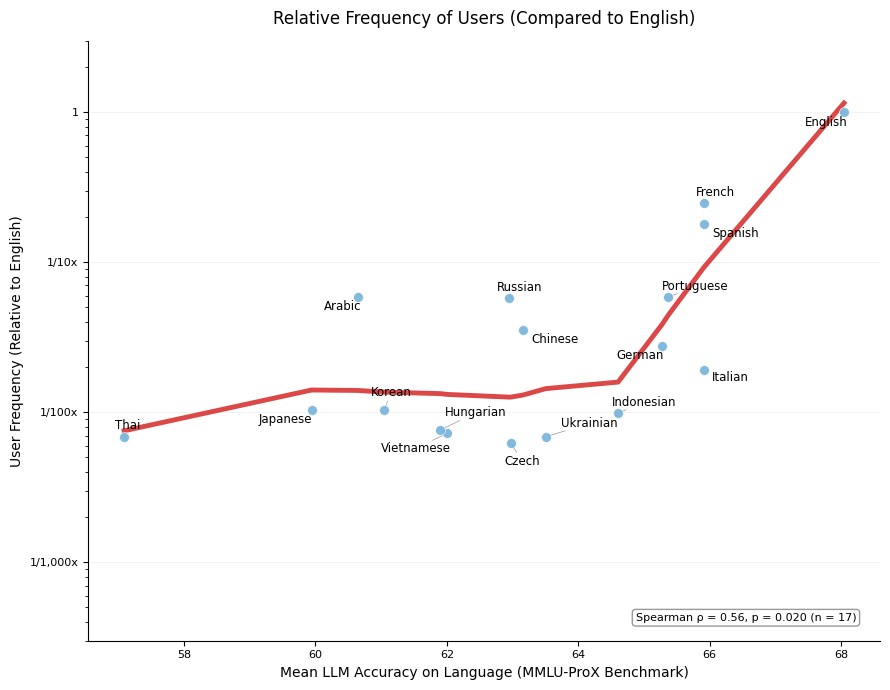}
\caption{Language Usage Frequency (relative to English) versus LLM Performance (average MMLU-ProX benchmark scores). The red curve is a LOWESS fit (frac = 0.6). Points are shown for languages used by $\geq$ 50 users and for which MMLU-Prox scores were available.}\label{fig6:mmlu}
\end{figure}

To examine whether model performance shapes users' choice of language, we examine the prevalence of different languages relative to their performance on the MMLU-proX benchmark. The usage of non-English languages tends to be flat and generally infrequent ($<$1/100th the frequency of English even for languages like Arabic that are spoken by large fractions of the world's population) until languages attain an MMLU score of approximately 65, at which point usage is higher and generally increasing as average scores increase (Figure~\ref{fig6:mmlu}). Notably, and consistent with the prior results in Figure~\ref{fig5:language}, European languages generally achieve higher scores; Asian languages have lower scores. African languages are effectively missing from usage entirely despite the large population of the African continent. Notably, African languages also score considerably lower, on average, on the MMLU-ProX benchmark (Table~\ref{tab:mmlu_regional}).

\begin{table}[t]
\centering
\caption{Mean MMLU-ProX accuracy across 15 LLMs by language and region. N indicates the count of languages scored by region (consistent with Figure~\ref{fig5:language}).}
\label{tab:mmlu_regional}
\small
\begin{tabular}{llr}
\toprule
\textbf{Region} & \textbf{N} & \textbf{MMLU-ProX} \\
\midrule
\textit{Americas \& Europe} & 12 & 64.6 \\
\textit{Asia \& Oceania} & 12 & 58.1 \\
\textit{Africa} & 5 & 41.3 \\
\bottomrule
\end{tabular}
\end{table}

\section{Discussion}

This study characterized differences in how early adopters used a major, free AI chatbot across countries around the world. We find two major results. First, we document \textit{higher} relative rates of usage for schooling and market work and \textit{lower} rates of usage for leisure in low- and middle- versus high-income countries. Second, we identify a gap in usage related to language, with usage globally disproportionately taking place in English. English is particularly likely to be overrepresented, we find, in countries where the dominant language performs poorly on model benchmarks. The gaps we observe in AI usage with respect to language should be of particular concern to model developers and policymakers seeking to enable global access and use.

The patterns we observe in domain of use run counter to prior research. Classic studies of second-level divides in internet usage, for example, find \textit{less} usage for productivity or learning and \textit{more} usage for distraction among historically marginalized versus advantaged groups \cite{goel2012does, van2020digital}. There are three possible explanations for the patterns we observe. First, high usage for schooling might be a product of user selection, or differences in \textit{who} is adopting generative AI across places. If students tend to be the earliest adopters of AI, for example, and low-income countries are earlier along the adoption curve relative to high-income countries, the over-representation of students would produce figures like those we observe. 
Second, our findings are likely shaped by user substitution between chatbots. \citet{chatterji_how_2025} for example, find that OpenAI's ChatGPT is used for writing; \citet{handa_which_2025} finds that Anthropic's Claude is most commonly used for programming and computer science; and Microsoft's Copilot is commonly used for knowledge work~\cite{tomlinson_working_2025}. 
People doing paid market work might have invested in paid model providers, while users in low-income countries may have been less able or willing to make those monthly payments. 
Finally, the patterns we observe might represent true underlying differences in the needs and applications for which users from different places sought out chatbots. 

The presence of true underlying differences in needs and use would be consistent with prior evidence of important cultural differences in the ways in which people from different places understand and use search \cite{lee2018ai}. Our findings could indicate a strong potential for leapfrogging benefits if generative AI genuinely helps expand and improve learning (and making it important for model providers to explicitly invest in design features that facilitate long-term learning benefits rather than cheating.) Understanding the factors driving observed differences in use will be key to ensuring that all places are able to access the benefits and mitigate the harms of this emerging technology, and thus further research is needed to disentangle the drivers of the patterns this study surfaces.

Our finding of a gap in usage across languages, particularly for places where low-resource languages are spoken, is consistent with the predictions of prior research benchmarking model performance across different languages \cite{xuan_mmlu-prox_2025, singhGlobalMMLUUnderstanding2025, huangBenchMAXComprehensiveMultilingual2025a, romanou2025include,vayani2025all} and evidence of reduced adoption in low-resource language countries \cite{misra2025ai}. Our study is novel, however, in empirically characterizing the effects of these differences in real-world chatbot conversations. We note that interpreting differences in usage across language as a skill-based barrier risks placing the onus for mitigation on the user, i.e. expecting more people to learn high-resource languages in order to access key technologies. We caution against such an interpretation, and point instead to the important achievements model developers are already making in improving model performance across new languages. Notably, \citet{chatterji_how_2025} show that AI's use has expanded dramatically in lower and middle income countries in the last year---a change in which improving multilingual capabilities likely played a large role \cite{microsoft2026globalai}. There is, moreover, an opportunity to leverage the roll-outs of improved models as a natural experiment for studying the extent to which improvements in usability in a given place's predominant language expands usage, and to investigate precisely what kind of usefulness it enables. This would be a natural extension of the methods and frameworks of our paper. However, differential performance on safety and alignment benchmarks across different languages raises the risk that, even as first-level divides in access and second-level divides in skills and use are mitigated, so-called ``third-level'' divides in who experiences harms from use become more likely. 

\subsection{Limitations}

Our study is subject to three major limitations. First, because we use a country-stratified sample, our results are not representative of overall use. However, our approach has the benefit of enabling comparison between countries around the world. Second, we limited our data set to early adopters, people who regularly used the chatbot between April and September of 2024. We focus only on users who had at least five separate interactions with the AI chatbot because we are interested in people who find value in the tools, as indicated by return usage--but it means that we excluded the large fraction of users who interact only a few times.  Moreover, models have improved since this research was conducted, particularly with respect to performance across languages, and thus there is need for further research to understand how these patterns of use have shifted with time. Finally, our study is descriptive, and we cannot make causal claims about the factors driving the patterns we observe. Nevertheless, as a large-scale examination of usage of a major chatbot that was freely available across over 200 countries around the world, we believe that our study will be of value to researchers, model developers, and policymakers seeking to foster globally inclusive and beneficial use of this major emerging technology.

\section{Conclusion}

Generative AI's rapid global adoption is evidence of its usefulness to people around the globe. Our work shows that its particular usefulness for learning applications is being leveraged everywhere, including by users in low-income countries. Whether this technology expands digital divides versus fosters leapfrogging may depend, our work suggests, on how well models perform across languages. In addition to tackling traditional barriers to inclusive technology adoption, such as infrastructure and education, then, policymakers and model developers should prioritize linguistic access.

\section{Acknowledgments} This work was considerably improved by comments and feedback from Sid Suri, Kiran Tomlinson, Sonia Jaffe, Scott Counts, Jacy Anthis, Jake Hofman, Dan Goldstein, and members of the computational social science working group at Microsoft Research. We are also grateful to Hiwot Tesfaye, Dakota Blagg, Seth Spielman, and Connie Hsueh for guidance. This work was made possible by engineering support from Kate Lytvynets, Amber Hoak, and David Tittsworth. 

\bigskip

\bibliography{main}

\section{Ethical Statement}
This study was reviewed and approved by the Microsoft Research Institutional Review Board. 
The primary data for this study are proprietary data, and the usage of anonymized, de-identified, and privacy-scrubbed data for research purposes is included in the terms of use \\

\onecolumn

\section{Appendix}\label{Appendix}

\subsection{Prompts}\label{appendix:prompts}

All prompts were sent to GPT-4o-mini-2024-07-18 with temperature~0. Double braces (e.g., \texttt{\{\{TargetLanguage\}\}}) indicate placeholders for conversation-specific values.

\subsection{Classification Prompts (Pipeline Step~4)}\label{app:prompts:classification}

We used a two-stage approach to classify the domain of user conversations.

\subsubsection{Stage~1: User Profile Prompt}\label{app:prompts:user}

We first classified the complete set of conversations for a given user (included via the placeholder \texttt{\{\{Conversations\}\}}) and prompted the classifier to provide a brief, descriptive user profile. Note that this is an intermediate step only, and only final and aggregated conversation-level classifications are used in the final analysis.

\begin{quote}
\begin{verbatim}
Consider the following set of conversations between a
single user and an AI.

{{Conversations}}

Your task is to summarize information about the user and
their conversations overall, to provide context for another
annotator who will be labeling the ConversationDomain that
each individual conversation belongs to. Specifically you
are asked to:
1) Summarize what we learn about the user
   (employment/student status, field/level, any
   job/school specifics).
2) Describe the types of conversations they have with AI,
   and provide guidelines for mapping their conversations
   to ConversationDomains.

The ConversationDomain labels are below.
ConversationDomain:
- MARKET_WORK: User seeking help with paid work, job
  search, or projects that improve their employability.
  Includes non-school projects that improve the user's
  earning potential (e.g., personal websites, game
  development), work at non-profits, and work as a
  teacher, school administrator, PhD student, or other
  educator.
- SCHOOLING: User seeking help with their own degree-based
  learning or assignment-like work. Includes conversations
  where user asks formal or structured questions as might
  appear on a school assignment.
- HOUSEHOLD_PRODUCTION: User seeking help with household
  tasks that are substitutable by market services (e.g.
  cooking, cleaning, childcare, home/car/personal device
  maintenance, personal finance, event planning, obtaining
  housing).
- PERSONAL_CARE: User seeking help with performing care of
  the self (e.g. health, mental health, nutrition, sleep,
  fitness, grooming, beauty).
- LEISURE: Conversations are being used for
  entertainment/hobbies/socialization/play/informal
  self-learning with no clear external stakes or deadline.
- UNCLEAR: Messages are uninterpretable or so
  generic/brief they could plausibly belong to any
  ConversationDomain (e.g., 'hello', other one-word
  prompts).

Tips:
- When a conversation spans multiple ConversationDomains,
  assign the ConversationDomain most strongly evidenced by
  the underlying conversations.
- Use UNCLEAR only if content of conversations is
  uninterpretable or so general they could plausibly
  belong to any ConversationDomain (e.g., 'hello').
- Market work can take many forms. Pay attention to
  whether the user's conversations, when taken as a
  whole, might be particularly useful for a specific type
  of market work.
- Learning does not indicate the user is a student. If
  the user is learning, try to infer the goal that they
  are pursuing.

Schema:
{Schema}

Provide your output in the following format:

<answer>
{...valid JSON only...}
</answer>
\end{verbatim}
\end{quote}

\subsubsection{Stage~1: User Profile Output Schema}\label{app:prompts:user_schema}

The \texttt{\{Schema\}} placeholder in the user profile prompt is replaced with the following JSON schema, which defines the two fields the model must return.

\begin{quote}
\begin{verbatim}
[
  {
    "name": "UserStatus",
    "dtype": "string",
    "description": "1 or more sentence describing what
      evidence is available concerning the user's
      occupation and/or educational status. Focuses on
      whether they are employed, unemployed, or a
      student; if they are employed, what their job is;
      if they are a student, their field of study and
      the level of education they are currently
      completing."
  },
  {
    "name": "DomainGuidelines",
    "dtype": "string",
    "description": "A set of guidelines for mapping the
      user's conversations to ConversationDomains.
      Describes how to identify and classify the
      different types of conversations that this
      specific user engages in."
  }
]
\end{verbatim}
\end{quote}

\subsubsection{Stage~2: Conversation Classification Prompt}\label{app:prompts:conversation}

We next applied a conversation-level classifier to produce the final classifications used in the paper. For each conversation, the placeholder \texttt{\{\{SourceConversation\}\}} contains the conversation text, and \texttt{\{\{User\}\}} and \texttt{\{\{DomainGuidelines\}\}} are the Stage~1 outputs for that user.

\begin{quote}
\begin{verbatim}
Consider the following conversation between a user and AI.

{{SourceConversation}}

Your role is to label the ConversationDomain of the
provided conversation according to the following
description.
ConversationDomain:
- MARKET_WORK: User seeking help with paid work, job
  search, or projects that improve their employability.
  Includes non-school project that improves the user's
  earning potential (e.g., personal websites, game
  development), work at non-profits, and work as a
  teacher, school administrator, PhD student, or other
  educator.
- SCHOOLING: User seeking help with their own
  degree-based learning or assignment-like work. Includes
  conversations where user asks formal or structured
  questions as might appear on a school assignment.
- HOUSEHOLD_PRODUCTION: User seeking help with household
  tasks that are substitutable by market services (e.g.
  cooking, cleaning, childcare, home/car/personal device
  maintenance, personal finance, event planning, obtaining
  housing).
- PERSONAL_CARE: User seeking help with performing care of
  the self (e.g. health, mental health, nutrition, sleep,
  fitness, grooming, beauty).
- LEISURE: Conversations are being used for
  entertainment/hobbies/socialization/play/informal
  self-learning with no clear external stakes or deadline.
- UNCLEAR: Messages are uninterpretable or so
  generic/brief they could plausibly belong to any
  ConversationDomain (e.g., 'hello', other one-word
  prompts).

Tips:
- When a conversation spans multiple ConversationDomains,
  assign the ConversationDomain most strongly evidenced by
  the underlying conversations.
- Use UNCLEAR only if content of conversations is
  uninterpretable or so general they could plausibly
  belong to any ConversationDomain (e.g., 'hello').
- Market work can take many forms. Pay attention to
  whether the user's conversations, when taken as a
  whole, might be particularly useful for a specific type
  of market work.
- Learning does not indicate the user is a student. If
  the user is learning, try to infer the goal that they
  are pursuing.

For context, another annotator has provided a description
of the user and guidelines for labeling their
conversations.
User:
    {{User}}
Labeling Guidelines:
    {{DomainGuidelines}}

Focus on the user's messages and intent, rather than the
messages/intent of the AI. If the user mixes tasks
belonging to multiple domains within this conversation,
classify it according to whichever set of tasks is most
frequent.
Before providing your response, please use the `quotes`
and `processing` tags to break down your thinking. Your
response should be in English. Provide your output in the
following format:

<quotes>
2 or more quotes from the user relevant to labeling task.
</quotes>

<processing>
2-3 sentences making sense of the conversation.
</processing>

<answer>
A single ConversationDomain. Does not include any
additional text beyond the name of the ConversationDomain.
</answer>

Using this information, please provide your response now.
\end{verbatim}
\end{quote}

\subsubsection{Stage~2: Conversation Domain Output Schema}\label{app:prompts:conversation_schema}

The conversation classification prompt uses structured output with the following JSON schema.

\begin{quote}
\begin{verbatim}
[
  {
    "name": "ConversationDomain",
    "dtype": "enum",
    "description": "The area of life in which the user
      is seeking assistance from AI.",
    "categories": {
      "MARKET_WORK": "The user is seeking support with
        paid work or job searching, including for work
        at non-profits, as a side hustle, or as a part
        of a PhD. Includes performing, learning to
        perform, or preparing for tasks for employment,
        clients, or compensation.",
      "SCHOOLING": "The user is seeking support with
        their own degree-based learning, for example
        performing homework. Includes asking questions
        with formal or structured language as might
        appear on a school assignment (e.g. multiple
        choice or fill-in-the-blank).",
      "HOUSEHOLD_PRODUCTION": "The user is seeking
        support with tasks for themselves or their
        household that could be substituted with market
        goods/services, for example cooking, electronic
        device maintenance, family event planning,
        organizing, meal planning, cleaning, car
        maintenance, home maintenance, childcare,
        personal finance, attaining housing, or caring
        for family members.",
      "PERSONAL_CARE": "The user is seeking support
        with care of the self, for example their
        health, mental health, nutrition, sleep,
        fitness, grooming, or beauty.",
      "LEISURE": "The user is performing leisure
        activities, for example entertainment, hobbies,
        informal self-learning, games, social chatter,
        creative play, creating random images.",
      "UNCLEAR": "The user's message(s) are
        uninterpretable, or so brief or general that
        they could belong to any category."
    }
  }
]
\end{verbatim}
\end{quote}

\subsection{Translation Prompts}\label{app:prompts:translation}

To facilitate labeling, conversations in the development and validation sets were translated into English in a two-step procedure following \citet{wu2025please}. We first normalized the data and then used the GlotLID-V3 model to detect language, translating any messages with a detected language other than English, a symbol ratio below 0.5, and more than two tokens (excluding any gibberish or symbol-only conversations). 

\subsubsection{First-Pass Translation Prompt}\label{app:prompts:translation1}

\begin{quote}
\begin{verbatim}
Please translate the following text to English.
Provide only one translation and do not output anything
else after that.

<source>{{NormalizedText}}</source>
\end{verbatim}
\end{quote}

\subsubsection{Second-Pass Translation Prompt}\label{app:prompts:translation2}

\begin{quote}
\begin{verbatim}
Please improve the following translation to
English, ensuring your translation uses both
the correct script and correct language. Provide only one
translation and do not output anything else after that.

<source>{{NormalizedText}}</source>

<draft>{{FirstDraftTranslatedText}}</draft>
\end{verbatim}
\end{quote}

\subsection{Classifier Validation}\label{appendix:validation}

We take a human-in-the-loop approach to validating our classifier \cite{pangakis_keeping_2025}. We collected additional conversations for a development and validation set, described in Table~\ref{tab:irr-descriptives}.

\begin{table}[h]
\centering
\caption{Descriptive Statistics: Development and Validation Set}
\label{tab:irr-descriptives}
\begin{tabular}{lrrr}
\toprule
                & Development & Validation & Combined \\
\midrule
Users           &          50 &         58 &      108 \\
Conversations   &         590 &        756 &    1,346 \\
Countries       &          40 &         51 &       82 \\
\midrule
\multicolumn{4}{l}{\textit{Users by income group}} \\
\quad High              &  13 &  13 &  26 \\
\quad Upper-middle      &  11 &  10 &  21 \\
\quad Lower-middle      &  13 &  19 &  32 \\
\quad Low               &  13 &  16 &  29 \\
\bottomrule
\end{tabular}
\end{table}

\subsubsection{Classifier Development}

We first constructed a development data set, for which two human raters
independently reviewed and labeled the conversations and then discussed and resolved disagreements. To protect user privacy, all data were privacy scrubbed before review, human labeling was limited to  internal reviewers approved for data access (the authors), and the conversations were deleted after the labeling and prompt development was completed.

We constructed our classifier iteratively using this development data set. We noted two important insights. First, we found a large increase in accuracy using a two-stage classifier in which the AI first, provided a concise summary of all conversations and, in a second pass, labeled conversations with the context of the summary.
This was consistent with the human raters' approach: reviewing multiple conversations often clarified the appropriate label for any individual conversation. 

Second, we found that GPT-4o and GPT-4o-mini performed comparably on the development set. 
Given the considerable cost- and time savings from using the smaller model, we used GPT-4o-mini for the validation and throughout the analysis.

\subsubsection{Inter-rater reliability}

We calculated inter-rater reliability (IRR) between the AI classifier
and two human coders (H1 and H2), who were blind to the output of the classifier
on the validation set. 
Again, all data were privacy-scrubbed; the labeling was conducted by the authors only; and user data were deleted after labeling was completed.

Table~\ref{tab:irr-kappa} reports Fleiss' $\kappa$, Cohen's $\kappa$,
and Krippendorff's~$\alpha$ with bootstrap 95\% confidence intervals
(1{,}000 iterations), as well as percent agreement, over six categories (the five categories described in Table~\ref{tab:ai_use_taxonomy} and an additional ``unclear'' category for gibberish or LLM classification errors).

\begin{table}[h]
\centering
\caption{Inter-rater Reliability}
\label{tab:irr-kappa}
\small
\begin{tabular}{l r@{~}l r@{~}l}
\toprule
 & \multicolumn{2}{c}{Development ($n\!=\!588$\footnotemark)} & \multicolumn{2}{c}{Validation ($n\!=\!756$)} \\
\cmidrule(lr){2-3}\cmidrule(lr){4-5}
Metric & $\hat\theta$ & 95\% CI & $\hat\theta$ & 95\% CI \\
\midrule
Fleiss $\kappa$ (H1, H2)           & 0.852 & [0.81, 0.89] & 0.605 & [0.56, 0.64] \\
Fleiss $\kappa$ (H1, H2, AI)       & 0.743 & [0.71, 0.77] & 0.603 & [0.57, 0.63] \\
Cohen $\kappa$ (AI vs.\ consensus) & 0.744 & [0.70, 0.79] & 0.743 & [0.70, 0.79] \\
Krippendorff $\alpha$ (H1, H2, AI) & 0.743 & [0.71, 0.77] & 0.603 & [0.57, 0.63] \\
\midrule
\% agree (H1, H2, AI)              & \multicolumn{2}{c}{73.3} & \multicolumn{2}{c}{56.2} \\
\% agree (AI vs.\ consensus)       & \multicolumn{2}{c}{81.9} & \multicolumn{2}{c}{80.5} \\
\bottomrule
\end{tabular}
\end{table}

We observe high reliability on the development set (Fleiss $\kappa$ = 0.852) versus just moderate to substantial reliability on the validation set (Fleiss $\kappa$ = 0.605). Reliability is higher and more consistent across both data sets on those conversations for which humans achieve agreement (Cohen's $\kappa$ = 0.744 development, 0.743 validation). Similarly, while all three reviewers agree on just 56.5\% of conversations overall, agreement is 80.5\% on those conversations where both humans agreed. Similarly, in pairwise comparisons, the AI--H1 pair achieves higher agreement than the
human--human pair on the validation set ($\kappa = 0.659$ vs.\ $0.609$, Table~\ref{tab:irr-heatmap}), confirming that the classifier is at least as reliable as a second human coder.

\begin{table}[h]
\centering
\caption{Pairwise Cohen's $\kappa$ and percent agreement (all six categories).}
\label{tab:irr-heatmap}
\small
\begin{tabular}{l*{3}{r}c*{3}{r}}
\toprule
 & \multicolumn{3}{c}{Development} & & \multicolumn{3}{c}{Validation} \\
\cmidrule(lr){2-4}\cmidrule(lr){6-8}
     & H1    & H2    & AI    & & H1    & H2    & AI    \\
\midrule
\multicolumn{8}{l}{\textit{Cohen's $\kappa$}} \\
H1   & ---   & 0.852 & 0.692 & & ---   & 0.609 & 0.659 \\
H2   & 0.852 & ---   & 0.688 & & 0.609 & ---   & 0.546 \\
AI   & 0.692 & 0.688 & ---   & & 0.659 & 0.546 & ---   \\
\midrule
\multicolumn{8}{l}{\textit{Percent agreement}} \\
H1   & ---   & 89.5  & 77.7  & & ---   & 69.8  & 73.8  \\
H2   & 89.5  & ---   & 77.6  & & 69.8  & ---   & 64.8  \\
AI   & 77.7  & 77.6  & ---   & & 73.8  & 64.8  & ---   \\
\bottomrule
\end{tabular}
\end{table}

\subsection{Additional Figures}\label{appendix:addlfigs}

\begin{figure*}[t]
\centering
\includegraphics[width=0.95\textwidth]{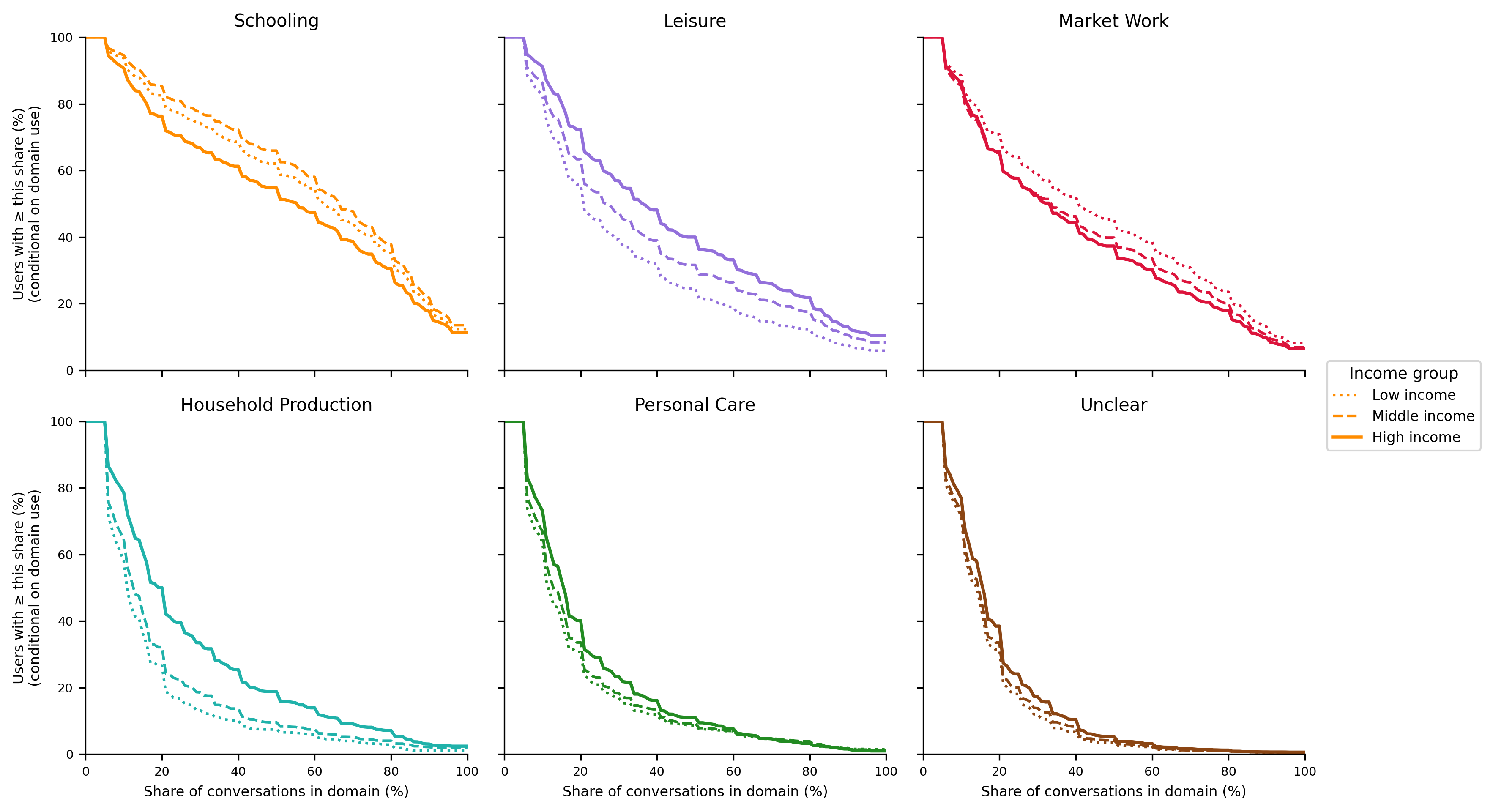}
\caption{Concentration of domain use by country-level income classification. Panels show the CCDF of the share of conversations devoted to each domain, conditional on users having engaged with that domain at least once, grouped by country-level income classification. Distributions are consistent across country income levels.}\label{appxfig2:cdf}
\end{figure*}

\end{document}